\documentclass[twocolumn,prl,showpacs]{revtex4} % galley
\usepackage{graphicx}

\newcommand{\be}{\begin{equation}}
\newcommand{\ee}{\end{equation}}
\newcommand{\barr}{\begin{eqnarray}}
\newcommand{\earr}{\end{eqnarray}}

\newcommand{\mgrader}{^\circ} % degrees in math and equations.
 
\newcommand{\Real}{\mathrm{Re}}

\newcommand{\unit}[1]{\mathrm {\,#1}}

\begin{document}
%\draft

\title{Unified treatment of fluorescence and Raman scattering processes
   near metal surfaces}

\author{Hongxing Xu, Xue-Hua Wang, Martin P. Persson, and H. Q. Xu}  

\affiliation{Division of Solid State Physics, 
   Lund University, Box 118, S--221\,00 Lund, Sweden}

\author{Mikael K\"all}

\affiliation{Department of Applied Physics, 
   Chalmers University of Technology, S--412\,96 G\"oteborg, Sweden}

\author{Peter Johansson}
\email[Corresponding author.]{}

\affiliation{Department of Natural Sciences, University of \"Orebro, 
   S--701\,82 \"Orebro, Sweden}

\date{\today}

\begin{abstract} 
We present a general model study of surface-enhanced resonant Raman
scattering and fluorescence focusing on the interplay between
electromagnetic effects and the molecular dynamics. Our model molecule
is placed close to two Ag nanoparticles, and has two electronic levels. 
A Franck-Condon mechanism provides electron-vibration coupling.  Using
realistic parameter values for the molecule we find that an
electromagnetic enhancement by 10 orders of magnitude can yield Raman
cross-sections $\sigma_{R}$ of the order $10^{-14} \unit{cm^2}$.
We also discuss the dependence of $\sigma_{R}$ on
incident laser intensity.
\end{abstract}


\maketitle

Discovered nearly three decades ago, surface-enhanced Raman scattering 
(SERS) has developed into an extremely sensitive spectroscopic technique 
with, in some cases, single molecule 
sensitivity\cite{Nie:97,Kneipp:97,Xu:99,Brus:99}. It is well known that 
SERS, as well as a range of related surface-enhanced optical processes, 
mainly results from electromagnetic (EM) effects
\cite{Reviews,Gersten:85}.
Although the Raman scattering cross-section $\sigma_R$ 
for a molecule in free space is very small (of the order of $10^{-30} 
\unit{cm}^2$ for nonresonant, $10^{-24} \unit{cm}^2$ for resonant 
scattering), the same molecule placed between two metal particles may 
well have an effective $\sigma_R$ that is 10--12 orders of magnitude 
larger.  The reason is that EM fields are strongly modified near and, in 
particular, between, metallic particles, so the local excitation field 
induced by an incident wave is much stronger there (by a factor that we 
denote $M$) than in free space. Likewise, by virtue of electromagnetic 
reciprocity, the amplitude of the radiation sent out from a source near 
the particles is equally enhanced compared with a source in free space. 
Consequently, quantities such as the absorption cross-section for a 
molecule increase by a factor $|M|^2$, whereas Raman scattering, which 
involves both an absorption and an emission event, increases by a factor 
$|M|^4$.

A large number of theory papers on surface-enhancement phenomena have 
focused on the electromagnetic aspects, but only a few have considered 
the molecular dynamics in more detail, see, for example, Ref.\ 
\onlinecite{Pettinger:86}. In this work we present a general model that 
treats the electromagnetic and molecular aspects on an equal footing. It 
includes photon-molecule coupling, coupling between electronic and 
vibrational degrees of freedom on the molecule, and radiative and 
non-radiative damping mechanisms\cite{CTnote}, 
and is analyzed by means of a density-matrix calculation.
The model lets us study not only how the molecule-metal-particle 
geometry affects
the EM enhancement and molecule damping rates, but also how these 
parameters in turn influence the spectrum of light emitted by the 
molecule. In particular, the model allows us to simultaneously quantify 
both scattering processes (Raman and Rayleigh) and fluorescence 
near metal surfaces, a field that has attracted a growing interest in 
recent years \cite{Barnes:98,Lakowicz:01}. By applying the model to the 
case of a highly fluorescent molecule situated between silver 
nanoparticles, we obtain an effective Raman cross-section 
of the same order of magnitude as in recent single-molecule SERS 
data\cite{Nie:97,Brus:99}.
In addition, we study the effects of a strong incident field, which drives
the molecule out of thermal equilibrium, and predict that it is possible 
to observe effects such as anti-Stokes Raman scattering even at low 
temperatures.

\begin{figure}[tb]
   \includegraphics[angle=0,width=7.5 cm]{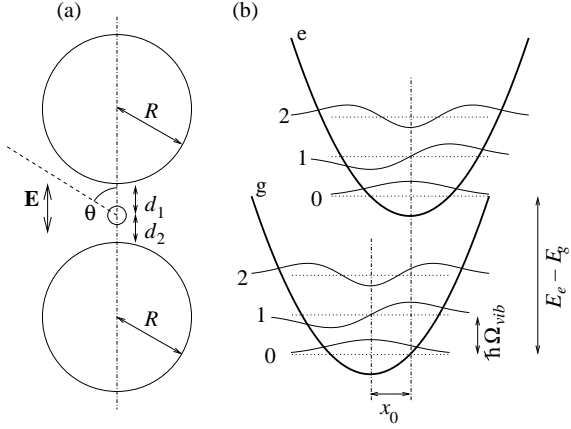}
   \caption{
      Schematic illustration of  (a) the model geometry and (b) the
      oscillator potential governing the vibrational motion of the
      molecule in the electronic ground and excited states.
   }
\label{fig:schematic}
\end{figure}

Figure \ref{fig:schematic} schematically shows the main ingredients
of the model.  A molecule is placed on the symmetry axis between two
spherical, metallic (Ag) nanoparticles. This system is illuminated from
the side ($\theta=90\mgrader$) by a laser with light polarized along the
symmetry axis.  The scattered and fluorescent light is collected by a
detector also placed on the side of the nanoparticle system. 

We treat the molecule as an electronic two-level system with an
excitation energy $E_e-E_g=\hbar\Omega_{ge}$.  The model also includes
one symmetric molecular vibration mode with frequency
$\Omega_{\mathrm{vib}}$ and reduced mass $\mu$.
The electronic and vibrational degrees of freedom are coupled by a
Franck-Condon mechanism.  As shown in Fig.\ \ref{fig:schematic} the
equilibrium position of the vibrational coordinate is displaced a
distance $x_0$ upon electronic excitation. 
The dimensionless
parameter $\alpha=x_0/\sqrt{2\hbar/(\mu\Omega_{\mathrm{vib}})}$ 
characterizes the strength of the electron-vibration coupling which
ultimately makes Raman processes possible.

The spectrum (differential cross-section per unit photon energy
$\hbar\omega$ and solid angle $\Omega$) of the light sent out by the
molecule in the direction of $\theta=90\mgrader$ as a
result of both scattering and fluorescence processes 
can be calculated from\cite{Scully}
\begin{equation}
   \frac{d^{\,2}\sigma}{d\Omega\,d(\hbar\omega)}
   =
   \frac{\omega^4|M(\omega)|^2}
   {I_{\mathrm{in}} 8\pi^3 c^3\varepsilon_0\hbar}
   \,\Real
   \int_{0}^{\infty} dt \, e^{i\omega t}
   \langle p^{(-)}(0) \, p^{(+)}(t) \rangle.
\end{equation}
Here  $I_{\mathrm{in}}$ is the incident laser intensity and $M(\omega)$
is the EM enhancement factor.  The normal-ordered correlation function
$\langle p^{(-)}(0) \, p^{(+)}(t) \rangle$ of the molecule dipole moment
$p$ can be evaluated by means of the quantum regression
theorem\cite{Scully} once we know the molecule's time-averaged density
matrix and its equation of motion. In this way light emitted both as a
result of molecular transitions (fluorescence) and as a result of the
molecule's oscillating dipole moment (scattering) is accounted for.  We
calculate the density matrix $\rho$ describing the molecule dynamics
keeping a finite number $N_{\mathrm{vib}}$ (usually 4) of vibrational
levels  per electronic level.  The equation of motion reads 
\begin{equation}
   i\, {d\rho}/{dt}
   =
   ({1}/{\hbar}) 
   [H_{mol} + H', \,\rho] 
   + 
   {\cal L}_{\mathrm{tr}} \rho 
   +
   {\cal L}_{\mathrm{ph}} \rho,
\label{eq:motion}
\end{equation}
where $H'$ describes the molecule-laser
interaction, ${\cal L}_{\mathrm{tr}}$ and ${\cal L}_{\mathrm{ph}}$
account for various damping processes which we specify below,
and $H_{mol}$ is the molecule Hamiltonian, 
\begin{equation}
   H_{mol}
   =
   \sum_{n=0}^{N_{\mathrm{vib}}-1}
   \sum_{l=g,e} 
      |l;n>(E_l+ n\hbar\Omega_{\mathrm{vib}}) <l;n|.
\end{equation}

The  molecule-electric-field  interaction is a central ingredient in the
dynamics that we account for by a $-e \vec{r}\cdot\vec{E}$ term in the
Hamiltonian.  The electric field $\vec{E}$ has contributions both from the
incident laser beam and the vacuum fluctuations. The laser field
\begin{equation}
   \vec{E}_L = 
   \hat{z} E_0 \cos{\Omega_L t} = 
   \hat{z} E_0\, [e^{i\Omega_L t} + e^{-i\Omega_L t}]/2,
\label{eq:laserfield}
\end{equation}
yields a transition matrix element (in the rotating-wave approximation)
between two molecule states  in the electronic ground and excited
states, respectively,
\begin{equation}
   \langle e;m | H' | g;n \rangle
   =
   -{M(\Omega_L)\, p_0E_0}
   \, e^{-i\Omega_L t} \, f(n,m)/{2}.
\end{equation}
Here 
$\hbar\Omega_L$ is the photon energy and
$p_0=e\ell_{\mathrm{dip}}$ is the transition dipole moment between the
electronic states. 
$f(n,m)$ is a Franck-Condon factor, i.e.\ the overlap
$f(n,m) = \langle 0;n|x_0;m \rangle$ between state $|n\rangle$ in the
undisplaced oscillator potential and state $|m\rangle$ in the displaced
potential, which depends on  the parameter
$\alpha=x_0/\sqrt{2\hbar/(\mu\Omega_{\mathrm{vib}})}$ 
introduced above\cite{FCnote}.

For a molecule in free space the interaction with the EM vacuum
fluctuations yield a decay rate from state $|e;m\rangle$ to
$|g;n\rangle$ due to spontaneous emission\cite{Sakurai},
\begin{equation} \Gamma_{gn, em} = {\omega^3} |p_0|^2 |f(n,m)|^2 \, / \,
({3\pi \hbar \varepsilon_0 c^3}) .  \end{equation} Near the metallic
particles the decay rate is modified, $\Gamma_{gn, em} \rightarrow |M_d
(\omega)|^2 \Gamma_{gn, em} $, where $\omega = \Omega_{ge} + (m-n)
\Omega_{\mathrm{vib}}$.  $|M_d|^2=P/P_{\mathrm{free}}$ is the ratio of
the  power emitted by a dipole placed at the position of the molecule
with and without the metal particles present.  Usually $|M_d|^2$ is of a
similar order of magnitude as $|M|^2$, yet the two factors may differ
substantially because $|M_d|^2$ accounts for radiation in all directions
as well as energy dissipation in the metal particles.  We calculate $M$
and $M_{d}$ using extended Mie theory\cite{Xu:00}.  The optical
properties of the particles are represented by a tabulated, local
dielectric function\cite{Palik}.  For small ($<$20--30 \r{A})
molecule-particle separations $d$,  there are important corrections
($\sim 1/d^4$) to the
damping-rate enhancement $|M_d|^2$ as a result of electron-hole pair
creation in the particles.  To capture this we calculate, in the
non-retarded limit while applying a long-wavelength cutoff, the power
$P_{\mathrm{eh}}$ dissipated by the dipole when placed between two flat
Ag samples (at the same distances as the spheres) whose optical
properties are described by a non-local dielectric function based on
d-parameter theory \cite{Liebsch:87}, and add this to the output power
$P_{\mathrm{Mie}}$ found in the Mie calculation, i.e.\
$|M_d|^2=(P_{\mathrm{Mie}}+P_{\mathrm{eh}})/P_{\mathrm{free}}$.

Decay and dephasing rates enter the last two terms of 
Eq.\ (\ref{eq:motion}).  
Standard quantum optics methods\cite{Scully} yield  
\begin{equation}
   {\cal L}_{\mathrm{tr}}\rho
   =
   -
   \sum_{kj} 
   \frac{i\Gamma_{kj}}{2} 
   \left[
      \sigma_{jk}\sigma_{kj}\rho + \rho \sigma_{jk} \sigma_{kj} 
      - 2 \sigma_{kj}\, \rho\, \sigma_{jk}
   \right]
\label{eq:liouville1}
\end{equation}
in the low-temperature limit.
($\sigma_{kj}$ denotes a matrix with the only non-zero element $kj$
equal to 1.) 
$\Gamma_{kj}$ is the total decay rate 
from state $j$ to $k$. It includes the radiative and non-radiative
processes
discussed above as well as vibrational damping due to transitions 
with a phenomenological rate $\gamma_{\mathrm{vib}}$ to the
nearest, lower level within the same electronic state. 
We also introduce a phenomenological dephasing rate 
$\gamma_{\mathrm{ph}}$ that enters the last term of Eq.\
(\ref{eq:motion}),
$
   {\cal L}_{\mathrm{ph}} \rho_{kj} 
   =
   - i \gamma_{\mathrm{ph}} \rho_{kj}
$
provided the electron states of $k$ and $j$ differ.
Let us stress that
$\gamma_{\mathrm{ph}}$ is brought into the model in order 
to broaden the fluorescence resonances of the molecule.
In reality an organic molecule has many vibration modes, and therefore
an almost continuous fluorescence spectrum. Dephasing gives us a
broadened fluorescence spectrum even though the model molecule has only
one vibrational mode. 
It has limited impact on the resonant Raman scattering as
long as $\gamma_{\mathrm{ph}}$ is smaller than, or comparable to, the
laser detuning.

\begin{figure}[tb]
   \includegraphics[angle=0,width=7.5 cm]{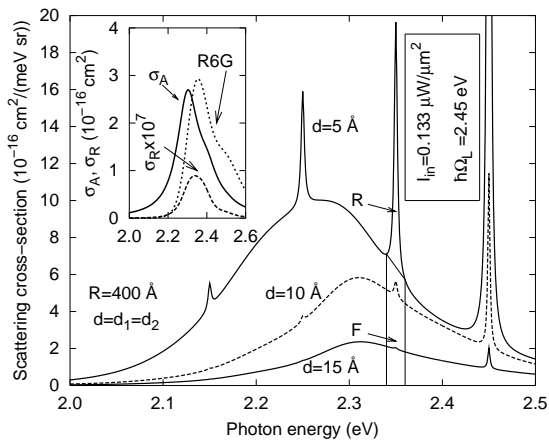}
   \caption{
      Combined Raman and fluorescence spectra 
      for a model molecule, with parameter values
      $\hbar\Omega_{ge}$=2.3 eV, 
      $\hbar\Omega_{\mathrm{vib}}$=0.1 eV, 
      $\gamma_{ph}$=10$^{14}$ s$^{-1}$, 
      $\gamma_{\mathrm{vib}}$=2$\times10^{12}$ s$^{-1}$, 
      $\alpha$=0.5,
      and
      $\ell_{\mathrm{dip}}$=1 \r{A},
      placed between two Ag particles with
      $R$=400~\r{A} for three different molecule-particle
      separations, and $\hbar\Omega_L$=2.45 eV.
      Inset:
      The absorption and Stokes Raman scattering cross-sections
      $\sigma_A$ and $\sigma_R$ as a function of the {\em incident}
      laser photon energy for the model molecule in {\em free space}.
      The experimental $\sigma_A$ for an R6G molecule is shown as a
      comparison.
   }
\label{fig:lettspect}
\end{figure}

The inset of Fig.\ \ref{fig:lettspect} shows the absorption
cross-section $\sigma_A$, and Raman profile ($\sigma_R$ as a function of
{\em incident} photon energy) calculated using the Fermi golden rule for
the model molecule in free space\cite{Sakurai}. The experimental
absorption cross-section for a Rhodamine 6G (R6G) dye molecule is shown
in the same diagram. We have set the parameter values cited in the
caption to obtain a similar spectrum. $\hbar\Omega_{ge}$ gives the
overall peak position and $\hbar\Omega_{\mathrm{vib}}$ the vibrational
quantum, and the value for $\gamma_{\mathrm{vib}}$ is reasonable for a
molecule at a metal surface. $\alpha$ has been chosen to reproduce the
shoulder of the spectrum, and $\gamma_{ph}$ and $\ell_{\mathrm{dip}}$
were set to reproduce the width and height of the R6G spectrum,
respectively. The free-molecule Raman cross-section is 7 to 8 orders of
magnitude smaller than $\sigma_A$, and its maximum is blue-shifted
compared with the absorption spectrum due to quantum-mechanical
interference between processes with different intermediate vibrational
states for the molecule.

The main panel in Fig.\ \ref{fig:lettspect} shows spectra calculated
with the model molecule placed between two silver spheres with radius
$R=400\unit{\r{A}}$ and three different, symmetric [i.e.\ $d_1=d_2$ in
Fig.\ \ref{fig:schematic}] molecule-particle separations. 
All three spectra have a broad fluorescence peak around
$\hbar\Omega_{ge}=2.3$ eV,  which
shifts to  slightly lower energy for $d$=5 \r{A} since the
maximum of $M$ is redshifted as the EM coupling between the Ag particles
increases. In addition 
a number of sharp peaks emerges, due to either Rayleigh scattering off the
molecule (at 2.45 eV) or Raman scattering (red-shifted by multiples of
$\hbar\Omega_{\mathrm{vib}}$=0.1 eV from 2.45 eV).  $\sigma_R$ varies
rapidly when the geometry is changed; for $d=5\unit{\r{A}}$ the Raman
peaks are comparable in height to the fluorescence background, whereas
for $d=15\unit{\r{A}}$ the Raman peak is barely discernible, 
cf.\ Ref.\ \onlinecite{Cotton:98}. 

The EM enhancement $|M|$ grows with decreasing $d$ yielding a rapid
growth of the Raman signal which involves both an absorption and
emission event and thus scales as $\sim|M|^4$.  Fluorescence also
involves both photon absorption and emission, but the cross-section in
this case only scales as $\sim |M|^4/|M_d|^2$.  The fluorescence
intensity is proportional to the EM enhancement in emission, $|M|^2$,
multiplied by the probability of finding the molecule in the excited
state. This probability is relatively insensitive to the enhancement
because it is set by the ratio ($\sim|M/M_d|^2$)  between the
laser excitation rate and the deexcitation rate due to
spontaneous emission. 

\begin{figure}[tb]
   \includegraphics[angle=0,width=7.0 cm]{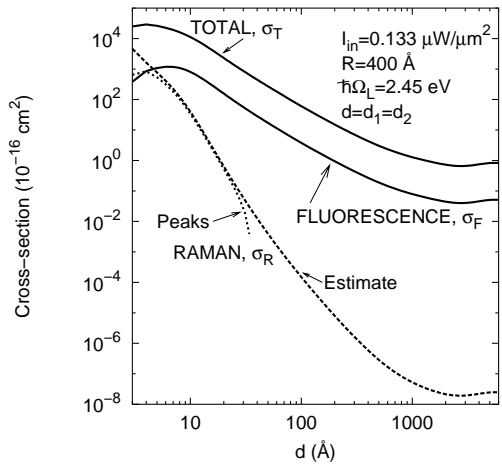}
   \caption{
      Calculated cross-sections as a function of molecule-metal-particle
      distance.
      The Raman (peaks) and fluorescence results have been extracted from the
      areas marked R and F, respectively, in Fig.\ \protect\ref{fig:lettspect}
      multiplied by the angular average $8\pi/3$. 
      The ``estimate'' curve gives the free molecule Raman cross-section
      multiplied by the relevant enhancement factor
      $|M(\Omega_L)|^2 |M(\Omega_L-\Omega_{\mathrm{vib}})|^2$.
      The ``total'' curve is the total cross-section integrated over the
      energy range 1.5--2.8~eV\@.
   }
\label{fig:distancedep}
\end{figure}

Figure \ref{fig:distancedep} shows the dependence  
of the Raman ($\sigma_R$), fluorescence ($\sigma_F$), and total
($\sigma_T$) cross-sections on the distance $d$,
exhibiting the same general tendencies as discussed above.
$\sigma_F$  has been calculated from the area marked F in Fig.\
\ref{fig:lettspect}.
For the Raman scattering we have plotted two curves: one is obtained
from the area marked R in Fig.\ \ref{fig:lettspect}, but this
calculation only works for relatively small $d$, so we also
estimate $\sigma_R$ by multiplying the free molecule Raman 
cross-section by the enhancement factor 
$|M(\Omega_L)|^2 |M(\Omega_L-\Omega_{\mathrm{vib}})|^2$.
The estimate agrees well with the ``peaks'' result for $d\approx$15--25
\r{A} and, of course, gives the true result for larger $d$. For small
$d$ the two results differ, and here the peaks result is the true one;
it includes effects on the Raman scattering of the strong energy
dissipation that are not included in the estimate.

At large $d$ the cross-sections approach those of a free molecule.  The
influence of the particles causes a weak interference phenomenon; there is
a minimum in $\sigma_R$ and $\sigma_F$ at $d\approx$3000 \r{A}. For
smaller $d$,  both $\sigma_R$ and $\sigma_F$ increase as a result of EM
enhancement. 
Over a range of distances the enhancement factors roughly scale as
$1/d$ meaning that $\sigma_F\sim1/d^2$ and $\sigma_R\sim1/d^4$.
At $d \alt 30$ \r{A}, more complicated behavior sets
in. Resonant enhancement with resonance frequencies that shift with changing
geometry occurs, and damping effects become important. These first
affect $\sigma_F$ which in spite of an increasing enhancement $M$ levels
off around $d=10$ \r{A} and eventually decreases, because the molecule is
usually deexcited through a non-radiative process. For the smallest
distances $\sigma_R$ becomes larger than $\sigma_F$, but it 
eventually decreases due to the strong dissipation which 
damps also the coherent oscillations of the molecular dipole moment.

\begin{figure}[tb]
   \includegraphics[angle=0,width=7.0 cm]{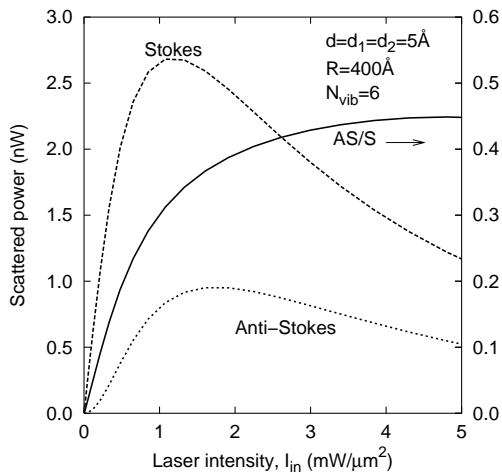}
   \caption{
      Stokes and anti-Stokes Raman scattering power (for
      low temperature) as
      well as their ratio as a function of $I_{\mathrm{in}}$  calculated
      with the parameter values given in Fig.\ \ref{fig:lettspect}.
   }
\label{fig:intensitydep}
\end{figure}

In Fig.\ \ref{fig:intensitydep} we show results for the power of both
Stokes and anti-Stokes Raman scattering as a function of incident laser
intensity $I_{\mathrm{in}}$. For low intensity, the Stokes power is linear 
in $I_{\mathrm{in}}$
(constant $\sigma_R$) while the anti-Stokes power grows quadratically
with $I_{\mathrm{in}}$. 
The anti-Stokes signal occurs because in an intense laser field the
molecule can be found in an excited vibrational level of the electronic
ground state once the rate of electronic excitation and deexcitation
becomes comparable to the vibrational damping rate
$\gamma_{\mathrm{vib}}$.
The probability for the molecule being vibrationally excited is 
roughly $C |M(\Omega_L)|^2 \sigma_A \phi_{ph}/\gamma_{\mathrm{vib}}$,
($C\alt1$ is a numerical factor),
and this ratio is
$\approx 0.6C$ with $|M(\Omega_L)|^2\approx1.4\times10^{5}$ and 
$I_{\mathrm{in}}$=0.5 mW/$\mu$m$^2$ corresponding to a photon flux 
$\phi_{ph}$=10$^{23}$ photons/(cm$^2$s).
Note that in this model the excited vibrational state is
pumped mainly by repeated absorption and deexcitation rather than by Raman
scattering\cite{Kneipp:98,Haslett:00}.
For higher intensities both the Stokes and anti-Stokes signals saturate
and eventually decrease. The molecule is driven so hard that its
polarizability becomes time-dependent. This happens when the effective
Rabi frequency $\Omega_R=M(\Omega_L)p_0E_0/\hbar$ becomes comparable 
to other relevant frequency scales,
in our case the dephasing rate $\gamma_{\mathrm{ph}}$.
For the parameter values used here 
$\Omega_R\approx 4\times10^{13}\unit{s}^{-1}$ 
at $I_{\mathrm{in}}=0.5\unit{mW}/\mu\unit{m}^2$.

In summary, we have presented a model calculation that treats
surface-enhanced Raman scattering and fluorescence on an equal footing.
We found that, with realistic parameter values, a resonant Raman 
cross-section of $\sim10^{-14}\unit{cm}^2$ can be reached with an EM
enhancement by 10 orders of magnitude. We also found that for an
incident laser intensity of $\sim 1 \unit{mW}/\mu\unit{m}^2$ it is
possible to get a considerable anti-Stokes Raman signal.

We acknowledge support from the Swedish Research Council and 
the Swedish Foundation for Strategic Research,  
and thank Lars Samuelson for useful discussions.

%%%%%


\begin{thebibliography}{99}
\bibitem{CTnote} Charge transfer processes between the metal particles
and the molecule, which may also play a role in SERS, are beyond the
scope of this model.
\bibitem{FCnote} The simplest Franck-Condon factors are given by 
$f(0,0)=e^{-\alpha^2/2}$, 
$f(1,1)=e^{-\alpha^2/2} (1-\alpha^2)$,
and 
$f(1,0)=-f(0,1)=\alpha\, e^{-\alpha^2/2} $.
For small $\alpha$, $\sigma_R\sim \alpha^2$.
\end{thebibliography}
\end{document}